\documentclass[twocolumn,aps,showpacs,prl]{revtex4}
\usepackage{graphicx}
\usepackage{epstopdf}
\usepackage{amsmath}
\usepackage{multirow}

\begin{document}

\title{Bi-collinear antiferromagnetic order in the tetragonal
$\alpha$-FeTe}

\author{Fengjie Ma$^{1,3}$}
\author{Wei Ji$^{3}$}
\author{Jiangping Hu$^{2}$}
\author{Zhong-Yi Lu$^{3}$}\email{zlu@ruc.edu.cn}
\author{Tao Xiang$^{4,1}$}\email{txiang@aphy.iphy.ac.cn}

\date{\today}

\affiliation{$^{1}$Institute of Theoretical Physics, Chinese Academy
of Sciences, Beijing 100190, China }

\affiliation{$^{2}$Department of Physics, Purdue University, West
Lafayette, Indiana 47907, USA}

\affiliation{$^{3}$Department of Physics, Renmin University of
China, Beijing 100872, China}

\affiliation{$^{4}$Institute of Physics, Chinese Academy of
Sciences, Beijing 100190, China }

\begin{abstract}

By the first-principles electronic structure calculations, we find
that the ground state of PbO-type tetragonal $\alpha$-FeTe is in a
bi-collinear antiferromagnetic state, in which the Fe local moments
($\sim2.5\mu_B$) are ordered ferromagnetically along a diagonal
direction and antiferromagnetically along the other diagonal
direction on the Fe square lattice. This bi-collinear order results
from the interplay among the nearest, next nearest, and next next
nearest neighbor superexchange interactions $J_1$, $J_2$, and $J_3$,
mediated by Te $5p$-band. In contrast, the ground state of
$\alpha$-FeSe is in the collinear antiferromagnetic order, similar
as in LaFeAsO and BaFe$_2$As$_2$.

\end{abstract}


\maketitle


The recent discovery of superconductivity in the layered iron-based
compounds attracts great interest, not only because the second
highest superconductivity temperatures are achieved, but also
because it is the first time that Fe atoms directly play an
important role in superconductivity while the Fe atoms are
conventionally considered in disfavor of superconductivity. Until
now there are four types of iron-based compounds that have been
reported to show superconductivity when doping or under high
pressures. They are classified as `1111' type (prototype: LaFeAsO
\cite{kamihara}), `122' type (prototype: BaFe$_2$As$_2$
\cite{rotter}), `111' type (prototype: LiFeAs \cite{wang}), and `11'
type (prototype: tetragonal $\alpha$-FeSe(Te) \cite{hsu,wu2}). The
common feature shared by these compounds is that there are the
robust tetrahedral layers where Fe atoms are tetragonally
coordinated by pnictide or chalcogenide atoms and the
superconduction pairing happens on these layers.

It has been found that for both `1111' and `122' compounds there
happens a tetragonal-orthorhombic structural phase transition
\cite{mook}. For undoped `1111', about 15K below this structural
transition, a magnetic phase transition takes place while for
undoped `122' compounds, the structural and magnetic phase
transitions occur at the same temperature. The magnetic order
associated with the transition is a collinear antiferromagnetic
order. The microscopic mechanisms underlying the structural
transition and antiferromagnetic transition  and the relationship
between these two transitions have been centered in  hot debate.
There are two very different microscopic pictures in understanding
the phase transitions. The first one suggests that there are no
local moments and the collinear antiferromagnetic order is entirely
induced by the Fermi surface nesting which is also likely
responsible for the structural transition due to breaking the
four-fold rotational symmetry\cite{Dong,Mazin}. The second one
suggests that As-bridged superexchange antiferromagnetic
interactions between the nearest neighbor and next nearest neighbor
Fe-Fe fluctuating local moments are the driving force upon the two
transitions \cite{ma,yildirim,Fang}.

In this letter, we report the electronic structure calculations on
teragonal $\alpha$-FeTe and $\alpha$-FeSe. We find that the ground
state of $\alpha$-FeTe is in a bi-collinear antiferromagnetic order
as schematically shown in Fig. 1(a), which differs from the
collinear antiferromagnetic order. The underlying physics of this
magnetic ordering can be described well in a model with the magnetic
exchange interactions $J_1$, $J_2$, and $J_3$ respectively between
the nearest, next nearest, and next next nearest neighbor mediated
by Te-$5p$ band (Fig. 1(a)). Hence, the fluctuating local moment
picture works well in understanding the magnetism while the magnetic
order can not be easily obtained from the Fermi Surface nesting
picture. In the bi-collinear antiferromagnetic state, the
bi-collinear antiferromagnetic ordering is along the diagonal
direction of the Fe-Fe square lattice with double collinearity. If
the Fe-Fe square lattice is divided into two square sublattices A
and B shown in Fig. 1(a), the Fe moments on each Fe-Fe sublattice
take its own collinear antiferromagnetic order. Our results are in
excellent agreement with neutron experimental results\cite{bao,Dai}.
In contrast, the ground state of tetragonal $\alpha$-FeSe is a
conventional collinear antiferromagnetic state, similar to LaFeAsO
and BaFe$_2$As$_2$.

\begin{figure}
\includegraphics[width=7cm]{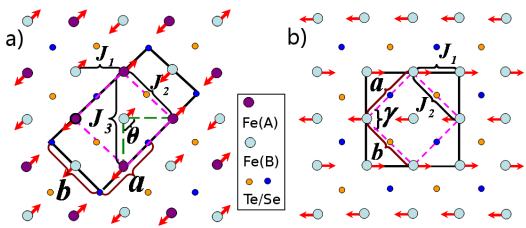}
\caption{(Color online) Schematic top view of the FeTe (FeSe) layer.
The ordered Fe spins are shown by red arrows. The small dashed
square is an $a\times b~(a=b)$ unit cell. (a) bi-collinear ordering
with a $2a\times b$ as the unit cell shown by solid rectangle; (b)
collinear ordering with a $\sqrt{2}a\times \sqrt{2} b$ as the unit
cell shown by solid square. } \label{figa}
\end{figure}


In the calculations the plane wave basis method was used
\cite{pwscf}. We employed the local (spin) density approximation and
the generalized gradient approximation of Perdew-Burke-Ernzerhof
\cite{pbe} for the exchange-correlation potentials. The ultrasoft
pseudopotentials \cite{vanderbilt} were used to model the
electron-ion interactions. After the full convergence test, the
kinetic energy cut-off and the charge density cut-off of the plane
wave basis were chosen to be 800eV and 6400eV, respectively. The
Gaussian broadening technique was used and a mesh of $18\times
18\times 9$ k-points were sampled for the Brillouin-zone
integration. In the calculations, the internal atomic coordinates
within the cell were determined by the energy minimization. Among
all the four types of iron-based compounds, $\alpha$-FeSe and
$\alpha$-FeTe possess the simplest structure, made up only of two
kinds of atoms Fe and Se or Te. They consist of continuous stacking
of tetrahedral FeSe or FeTe layers along c-axis without any other
kind of layers. $\alpha$-FeSe and $\alpha$-FeTe belong to a
tetragonal family with $PbO$-type structure and space group of
$P4/nmm$. The experimental tetragonal lattice parameters were
adopted in our calculations \cite{hsu,wu2,bao,Dai,okamoto}.

Our nonmagnetic calculations on both $\alpha$-FeSe and $\alpha$-FeTe
exclude any possible structural distortions like Jahn-Teller effect.
This means that a structural distortion on $\alpha$-FeSe or
$\alpha$-FeTe if existing is very likely driven by magnetic forces.
The calculated nonmagnetic electronic energy band structures and
Fermi surfaces are the same as those reported in Ref. \cite{singh}.

Now let's consider possible magnetic orders. If we divide the Fe-Fe
square lattice into two square sublattices $A$ and $B$, there may be
ferromagnetic, square antiferromagnetic, or collinear
antiferromagnetic order for the Fe moments on each of the two
sublattices. The combination of these magnetic orders on the two
sublattices yield the ferromagnetic, square antiferromagnetic,
collinear antiferromagnetic, and bi-collinear antiferromagnetic
orders on the Fe-Fe square lattice, as shown in Fig. 1. To calculate
the bi-collinear antiferromagnetic order and collinear
antiferomagnetic order, we adopt $2a\times a\times c$ and
$\sqrt{2}a\times\sqrt{2}a\times c$ unit cells, respectively shown in
Fig. 1(a) and (b). For the other magnetic orders, the $a\times
a\times c$ unit cell is used.

If the energy of the nonmagnetic state is set to zero, through the
calculations we find that the energies of the ferromagnetic
($E_{FE}$), square antiferromagnetic ($E_{AFM}$), collinear
antiferromagnetic ($E_{COL}$), and bi-collinear antiferromagnetic
states ($E_{BI}$) are (0.183, -0.101, -0.0152, -0.089) eV/Fe for
$\alpha$-FeSe and (-0.0897, -0.0980, -0.156, -0.166) eV/Fe for
$\alpha$-FeTe, respectively. Thus the ground state of $\alpha$-FeSe
is a collinear-ordered antiferromagnetic state, similar to LaFeAsO
and BaFe$_2$As$_2$. In contrast, the ground stat of $\alpha$-FeTe is
a bi-collinear ordered antiferromagnetic state, being lower in
energy by $\sim$10 meV/Fe than its collinear ordered
antiferromagnetic state.

The magnetic moment located around each Fe atom is found to be about
$2.2\sim 2.6~\mu_B$, similar to the cases of LaFeAsO and
BaFe$_2$As$_2$, varying weakly in the above four magnetically
ordered states. Since these local moments are embedded in the
environment of itinerant electrons, the moment of Fe ions is
fluctuating. Nevertheless, we may here reasonably consider the spin
of Fe ions is equal to 1. But the corresponding long range ordered
moment should be smaller because the correlated effect, especially
the strong competition between different antiferromagnetic states,
has not been fully included in the density functional theory
calculation.

To quantify the magnetic interactions, we assume that the energy
differences between these magnetic orderings are predominantly
contributed from the interactions between the Fe spins ($S=1$),
which can be modeled by the following frustrated Heisenberg model
with the nearest, next-nearest, and next next nearest neighbor
couplings $J_1$, $J_2$, and $J_3$,
\begin{equation}\label{eq:Heisenberg}
H=J_1\sum_{\langle ij \rangle}\vec{S}_i\cdot\vec{S}_j +J_2\sum_{ \ll
ij \gg}\vec{S}_i\cdot\vec{S}_j +J_3\sum_{ \langle\langle\langle ij
\rangle\rangle\rangle}\vec{S}_i\cdot\vec{S}_j,
\end{equation}
whereas $\langle ij \rangle$, $\ll ij \gg$ and
$\langle\langle\langle ij \rangle\rangle\rangle$ denote the
summation over the nearest, the next-nearest and the
next-next-nearest neighbors, respectively. Since both $\alpha$-FeSe
and FeTe are semimetals, the above modeling may miss the certain
contribution from the itinerant electrons. However we believe that
the above simply modeling captures the substantial physics on the
magnetic structures. From the calculated energy data, we find that
for $\alpha$-FeTe $J_1=2.1 meV/S^2$, $J_2=15.8 meV/S^2$, and
$J_3=10.1 meV/S^2$ while for $\alpha$-FeSe $J_1=71 meV/S^2$, $J_2=48
meV/S^2$, and $J_3=8.5 meV/S^2$. Notice that the ferromagnetic state
is quite low in energy in $\alpha$-FeTe, which is reason why $J_1$
is so small.

It is well-known that for a $J_1$-$J_2$ antiferromagnetic square
lattice the frustration between $J_1$ and $J_2$ will destruct the
Neel state and induce a collinear antiferromagnetic order when
$J_2>J_1/2$ as shown in Fig. 1(b), in which we notice that each site
spin has two ferromagnetic connection neighbors and two
antiferromagnetic connection neighbors. Further inspection of Fig. 1
shows that $J_2$ and $J_3$ are the nearest and the next-nearest
couplings for each of sublattice A and B whereas $J_1$ is the
connection coupling between A and B. When $J_3>J_2/2$ and
$2J_2+4J_3>J_1$\cite{Ferrer}, a collinear antiferromagnetic ordering
will take place on each sublattice. It turns out that the
bi-collinear antiferromagnetic ordering occurs on the square lattice
as shown in Fig. 1(a), in which we notice again that each site spin
has two ferromagnetic connection neighbors and two antiferromagnetic
connection neighbors that doesn't cost more energy on $J_1$ in
comparison with the collinear antiferromagnetic order. Overall, when
$J_3>J_2/2$ and $J_2>J_1/2$, by recursion we may thus expect that a
bi-collinear antiferromagnetic state would be the ground state of
the frustrated $J_1$-$J_2$-$J_3$ Heisenberg model (1), which is in
good agreement with the derived $J_1$, $J_2$, and $J_3$ on
$\alpha$-FeTe and $\alpha$-FeSe. Especially, we can show that
$J_3=J_2/2+(E_{COL}-E_{BI})/4S^2$ and
$J_2=J_1/2+(E_{AFM}-E_{COL})/4S^2$ \cite{ma3}. The energy ordering
is $E_{FM}>E_{AFM}>E_{COL}>E_{BI}$ for $\alpha$-FeTe and
$E_{FM}>E_{BI}>E_{AFM}>E_{COL}$ for $\alpha$-FeSe. By model (1) we
may thus easily understand the complex magnetic structures displayed
by $\alpha$-FeTe and $\alpha$-FeSe.

When we further consider the possible spin-phonon interaction, it is
expected that there would be further lattice relaxation, similar to
spin-Peierls distortion, that is, being slightly longer along spin
anti-parallel alignment to further favor antiferromagnetic energy
and shorter along spin parallel alignment to further favor
ferromagnetic energy. Correspondingly, $\theta$ and $\gamma$
increase slightly respectively for the bi-collinear and collinear
cases (Fig. 1(a) and (b)). As a matter of fact, we indeed find such
extra small structural distortions for both $\alpha$-FeTe and
$\alpha$-FeSe. For $\alpha$-FeTe, $\theta$ increases to
92.03$^{\circ}$ with an extra energy gain of $\sim$5 meV/Fe while
for $\alpha$-FeSe, $\gamma$ increases to 90.5$^{\circ}$ with an
extra energy gain of $\sim$2 meV/Fe. As results, the crystal unit
cell of $\alpha$-FeTe ($\alpha$-FeSe) on FeTe (FeSe) layer deforms
from a square to a rectangle (rhombus), as shown in Fig. 1. However,
we find that this small lattice distortion affects weakly the
electronic band structure and the Fe moments.

Our calculations also show that for both $\alpha$-FeTe and
$\alpha$-FeSe the Fe magnetic moments between the nearest neighbor
layers FeTe (FeSe) prefer the anti-parallel alignment but with a
small energy gain less than 1 meV/Fe, similar to LaFeAsO. It is thus
very likely that here the magnetic phase transition would happen
below the structural transition temperature.

\begin{figure}
\includegraphics[width=8cm]{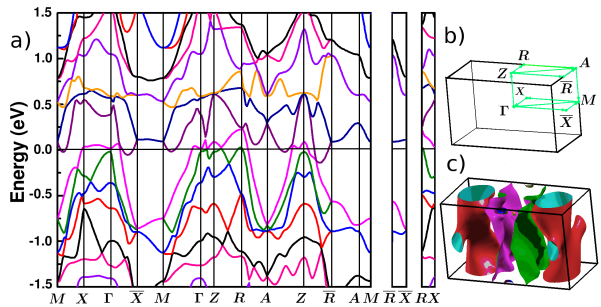}
\caption{(Color online) (a) Electronic band structure of the
bi-collinear-ordered antiferromagnetic $\alpha$-FeTe. The Fermi
energy is set to zero. (b) Brillouin zone. (c) Fermi surface.
$\Gamma \bar{X}$ ($\Gamma X$) corresponds to the
parallel(antiparallel)-aligned moment line.} \label{fig:2}
\end{figure}

Fig. \ref{fig:2} shows the electronic band structure and the Fermi
surface of $\alpha$-FeTe in the bi-collinear antiferromagnetic
order. There are three bands crossing the Fermi level which form
three discrete parts of the Fermi surface. The Fermi surface
contains a small hole-type pocket around R, two pieces of opened
irregular hole-type sheets parallel to the plane $\Gamma$-Z-R-X, and
two `blood vessel'-like electron-type cylinders between $\Gamma$ and
$\Bar{X}$. From the volumes enclosed by these Fermi surface sheets,
we find that the electron (hole) carrier density is 0.43
electron/cell (0.41 hole/cell), namely, $2.38\times 10^{21}/cm^3$
($2.26\times 10^{21}/cm^3$). In comparison with the nonmagnetic
state, the carrier density is just reduced by half. The density of
states at E$_F$ is 1.98 state/(eV$~$Fe), which yields the electronic
specific heat coefficients $\gamma$ = 4.665 $mJ/(K^2*mol)$.

\begin{figure}
\includegraphics[width=8cm]{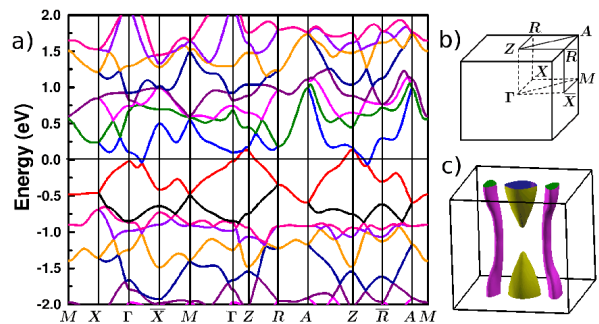}
\caption{(Color online) (a) Electronic band structure of the
collinear-ordered antiferromagnetic $\alpha$-FeSe. The Fermi energy
is set to zero. (b) Brillouin zone. (c) Fermi surface. $\Gamma X$
($\Gamma \bar{X}$) corresponds to the parallel(antiparallel)-aligned
moment line.} \label{fig:3}
\end{figure}

The electronic band structure and Fermi surface of $\alpha$-FeSe in
collinear antiferromagnetic state is shown in Fig. \ref{fig:3}. The
Fermi surface consists of three sheets formed by two bands crossing
the Fermi level. Specifically, one hole-type close sheet forms a
hole pocket centered around Z and the other two electron-type
cylinders are formed between $\Gamma$ and $\Bar{X}$, crossing the
Fermi level. The volumes enclosed by these sheets give the electron
(hole) carrier density 0.042 electrons/cell (0.028 holes/cell),
namely $2.68\times 10^{20}/cm^3$ ($1.81\times 10^{20}/cm^3$). In
comparison with the nonmagnetic state, the carrier density is
reduced by one magnitude while in LaFeAsO the corresponding
reduction is more than hundred times. The density of states at E$_F$
is 0.48 state/(eV~Fe), which yields the electronic specific heat
coefficients $\gamma$ = 1.129 $mJ/(K^2*mol)$.

\begin{figure}
\includegraphics[width=5cm]{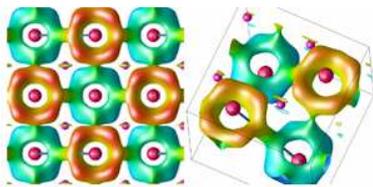}
\caption{(Color online) Top (left) and off-top (right) views of
differential charge density for $\alpha$-FeSe. The isosurface value
was set to 0.03 {\it e}/ \AA$^{3}$. Colors mapped on isosurfaces
represent the relative height of data points in $c$-direction.}
\label{fig:4}
\end{figure}

By projecting the density of states onto the five 3d orbitals of Fe,
we find that the five up-spin orbitals are almost completely filled
while the five down-spin orbitals are nearly uniformly filled by
half. This indicates that the crystal field splitting imposed by Se
or Te atoms is very weak and the Fe 3d-orbitals hybridize strongly
with each other. As the Hund rule coupling is strong, this would
lead to a large magnetic moment formed around each Fe atom, as found
in our calculations. Moreover, Fig. \ref{fig:4} clearly shows that
the band states constituted by Fe 3d orbitals are very localized.
Fig. \ref{fig:4} plots the differential charge density distribution
for $\alpha$-FeSe ($\alpha$-FeTe and LaFeAsO all have similar
distribution\cite{ma3}). We find that the most charge accumulations
are surrounding Se (Te, As) atoms (larger pink balls), while the Fe
atoms (smaller red balls) only have very small charge accumulation.
Figure \ref{fig:4} (right) further shows a differential charge
density pipeline connecting from a higher Se to a lower Se. For each
Se atom, there are four charge accumulation pipelines connecting
from it to its four adjacent Se atoms, suggesting an electron
network mostly formed by delocalized electrons of Se through
covalent bonding, in which the very localized Fe electron states are
embedded.

\begin{figure}
\includegraphics[width=7cm,height=5cm]{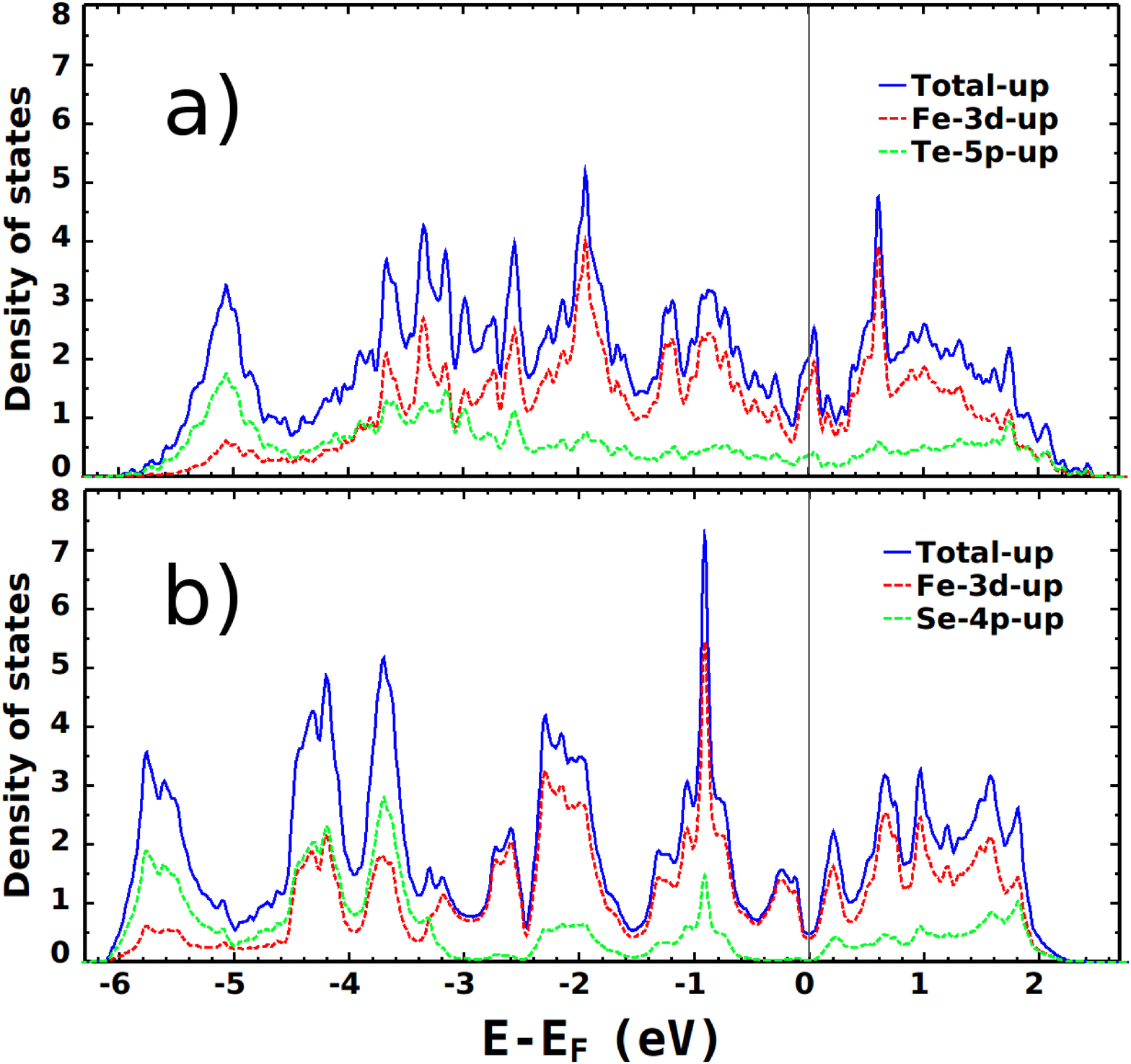}
\caption{(Color online) Total and orbital-resolved partial density
of states (spin-up part): a) bi-collinear antiferromagnetic
$\alpha$-FeTe; b) collinear antiferromagnetic $\alpha$-FeSe. }
\label{fig:5}
\end{figure}

On the other hand, Fig. \ref{fig:5} shows that the band formed by Se
4p orbitals (also As 4p orbitals) is gapped at the Fermi energy
while the band formed by Te 5p orbitals is partially filled. So
there are itinerant 5p electrons at Fermi energy involved in
mediating the exchange interactions in $\alpha$-FeTe, likely, a
long-range effect. This may explain why the coupling $J_3$ is nearly
zero for $\alpha$-FeSe, LaFeAsO and BaFe$_2$As$_2$.

It was reported\cite{hsu,wu2} in experiment that at 105K the
tetragonal $\alpha$-FeSe compound experiences a similar structural
distortion as LaFeAsO, with $a=b$ and $\gamma$ changing from
90$^{\circ}$ to 90.3$^{\circ}$ while at 45K the tetragonal
$\alpha$-FeTe compound also experiences a structural distortion but
with $a\neq b$ (a = 3.824 \AA, b = 3.854 \AA) and
$\gamma$=90$^{\circ}$. This is in excellent agreement with our
calculations. Furthermore, by the neutron scattering Bao and
coworkers \cite{bao} found that an incommensurate antiferromagnetic
order propagates along the diagonal direction of the Fe-Fe square
lattice, but they also find this incommensurate ordering is easily
tunable with composition and locks into a commensurate order in the
metallic phase. By our calculation, it is clear that the magnetic
order must be along the diagonal direction for $\alpha$-FeTe. A
physical picture suggested in our study is that the Fe moments are
mediated by Te $5p$-band and the origin $J_3$ exchange coupling may
be well induced through a RKKY-type mechanism. It is very likely
that the excess interstitial Fe moments drive the bi-collinear order
into the incommensurate order. Notice that this diagonal
antiferromagnetic order cannot be straightforwardly understood by
the Fermi surface nesting picture.

In conclusion, we have presented the electronic band structure of
$\alpha$-FeTe based on the first-principles density functional
theory calculations. Our studies show that the compound
$\alpha$-FeTe is a quasi-2-dimensional bi-collinear
antiferromagnetic semimetal with a magnetic moment of $\sim2.5\mu_B$
around each Fe atoms.


This work is partially supported by National Natural Science
Foundation of China and by National Program for Basic Research of
MOST, China. JPH were supported by the NSF of US under grant No.
PHY-0603759. We would like to thank P. Dai and S. Li for sharing his
unpublished neutron scattering results.

\end{document}